\documentclass[sigconf, nonacm]{acmart}
\AtBeginDocument{%
	\providecommand\BibTeX{{%
			\normalfont B\kern-0.5em{\scshape i\kern-0.25em b}\kern-0.8em\TeX}}}

\usepackage{multirow}
\usepackage{soul}
\usepackage{mathtools}
\usepackage{amsmath,amsfonts, amsthm}
\usepackage{graphicx}
\usepackage{float}
\usepackage{lipsum}
\usepackage{subfig}
\usepackage{textcomp}
\usepackage{xcolor}
\usepackage{placeins}
\usepackage{hhline}
\usepackage{tabularx}
\usepackage{enumitem}
\usepackage{setspace}
\usepackage{booktabs}
\usepackage{adjustbox}
\usepackage[normalem]{ulem}
\usepackage{indentfirst}
\usepackage{makecell}
\usepackage{titlesec}
\usepackage[T1]{fontenc}
\usepackage{textcomp, libertine}

\usepackage{algorithm}

\usepackage[]{algpseudocode}

\makeatletter
\newcommand{\mathleft}{\@fleqntrue\@mathmargin0pt}
\newcommand{\mathcenter}{\@fleqnfalse}
\makeatother

\theoremstyle{definition}
\newtheorem{problem}{\textbf{Problem}}

\let\sigproof\proof\let\proof\relax
\let\sigendproof\endproof\let\endproof\relax
\let\proof\sigproof
\let\endproof\sigendproof

\DeclareMathOperator*{\argmax}{arg\,max}

\makeatletter
\def\@copyrightspace{\relax}
\makeatother

\begin{document}
	\title{Source Localization for Cross Network Information Diffusion}
	
    \author{Chen Ling}
    \authornotemark[1]
    \affiliation{%
    \institution{Emory University}
    \city{Atlanta}
    \country{United States}}
    \email{chen.ling@emory.edu}
    
    \author{Tanmoy Chowdhury}
    \authornotemark[1]
    \affiliation{%
    \institution{Emory University}
    \city{Atlanta}
    \country{United States}}
    \email{tchowdh6@gmu.edu}       

    \author{Jie Ji}
    \affiliation{%
    \institution{Emory University}
    \city{Atlanta}
    \country{United States}}
    \email{jason.ji@emory.edu}

    \author{Sirui Li}
    \affiliation{%
    \institution{Emory University}
    \city{Atlanta}
    \country{United States}}
    \email{sirui.li@emory.edu}
    
    \author{Andreas Z\"ufle}
    \affiliation{%
    \institution{Emory University}
    \city{Atlanta}
    \country{United States}}
    \email{azufle@emory.edu}
    
    \author{Liang Zhao}
    \affiliation{%
    \institution{Emory University}
    \city{Atlanta}
    \country{United States}}
    \email{liang.zhao@emory.edu}

	
	\begin{abstract}
    Source localization aims to locate information diffusion sources only given the diffusion observation, which has attracted extensive attention in the past few years. Existing methods are mostly tailored for single networks and may not be generalized to handle more complex networks like cross-networks. Cross-network is defined as two interconnected networks, where one network's functionality depends on the other. Source localization on cross-networks entails locating diffusion sources on the source network by only giving the diffused observation in the target network. The task is challenging due to challenges including: 1) diffusion sources distribution modeling; 2) jointly considering both static and dynamic node features; and 3) heterogeneous diffusion patterns learning. In this work, we propose a novel method, namely \textit{CNSL}, to handle the three primary challenges. Specifically, we propose to learn the distribution of diffusion sources through Bayesian inference and leverage disentangled encoders to separately learn static and dynamic node features. The learning objective is coupled with the cross-network information propagation estimation model to make the inference of diffusion sources considering the overall diffusion process. Additionally, we also provide two novel cross-network datasets collected by ourselves. Extensive experiments are conducted on both datasets to demonstrate the effectiveness of \textit{CNSL} in handling the source localization on cross-networks.
    The code are available at: \href{https://github.com/tanmoysr/CNSL/}{https://github.com/tanmoysr/CNSL/}
	\end{abstract}

	
	\maketitle

    \vspace{-0.5cm}
    
    \setlength{\abovedisplayskip}{5pt} \setlength{\belowdisplayskip}{5pt}
    \section{Introduction}
    
    Source localization aims at locating the origins of information diffusion within networks, which stands as a famous inverse problem to the estimation of information propagation. Source localization not only holds practical significance but also helps us grasp the intricate characteristics of network dynamics. By accurately identifying the sources of information propagation, we can significantly mitigate potential damages by cutting off critical pathways through which information, and potentially misinformation, spreads. Over the past years, existing works have made considerable efforts toward addressing this critical challenge. Earlier works \cite{prakash2012spotting, zang2015locating, wang2017multiple, zhu2017catch} leverage \textit{rule-based} methods to locate diffusion sources under prescribed and known diffusion patterns. Furthermore, \textit{Learning-based} methods \cite{dong2019multiple, ling2022source, wang2022invertible,wang2023lightweight} were proposed to employ deep neural networks to encode neighborhood and graph topology information, which achieve state-of-the-art performance. These efforts underscore the importance of source localization in maintaining the integrity and reliability of information across the digital landscape.

    \begin{figure}
        \centering
        \includegraphics[width=0.81\columnwidth]{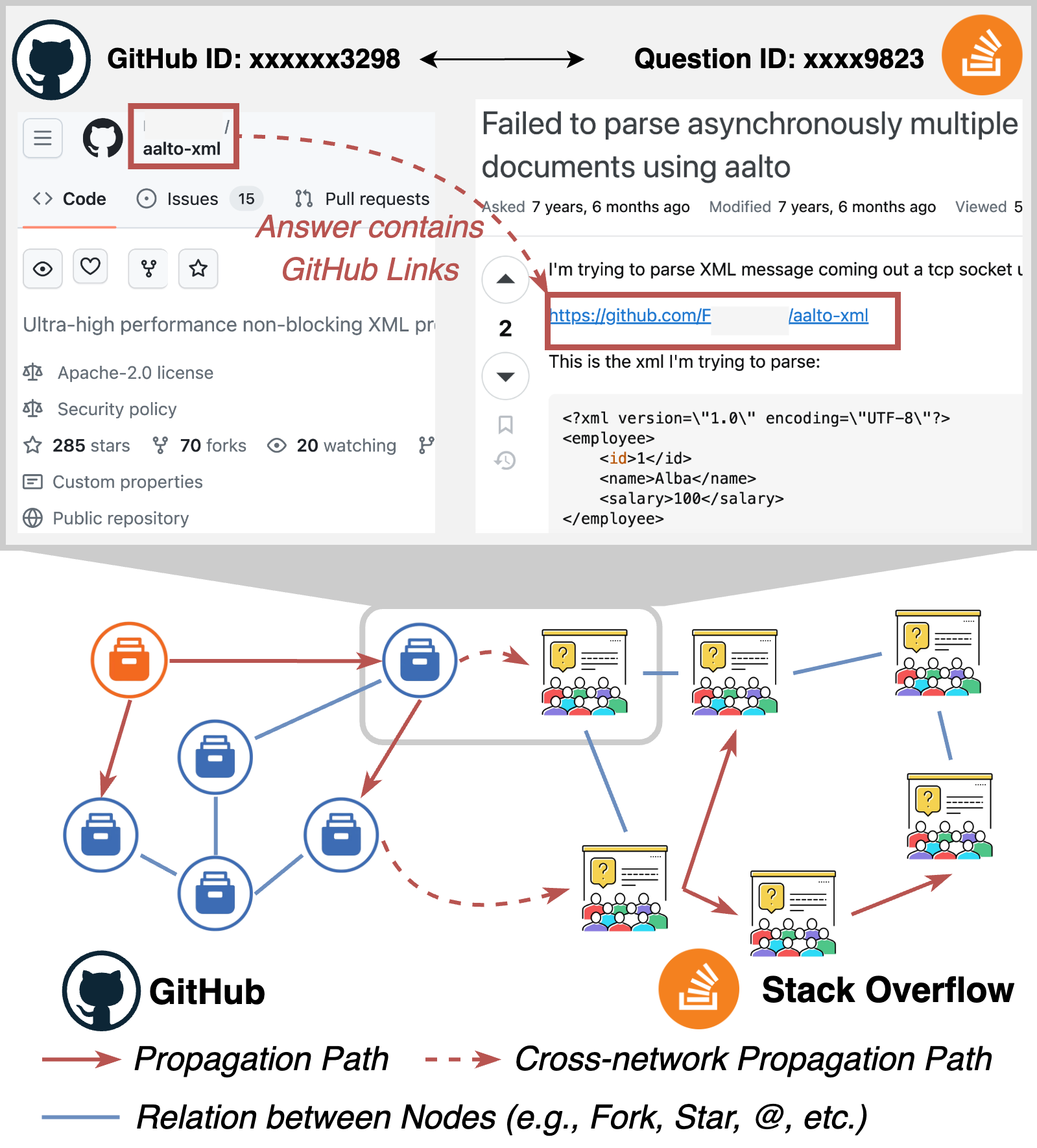}
        \vspace{-4mm}
        \caption{Example of misinformation propagation on cross-network between GitHub and Stack Overflow, where each node in the GitHub network denotes a repository, and each node in the Stack Overflow represents a discussion thread.}
        \label{fig: intro}
        \vspace{-7mm}
    \end{figure}

    Existing techniques for source localization have primarily been designed for single networks. However, much of today’s infrastructure is organized in the form of \textit{cross-networks}. Communications between different communities, cross-country financial transactions, and systems of water and food supply can all be cross-networks, where the functionality or performance of one network depends on other networks. The presence of cross-networks has also made us vulnerable to various network risks that belong exclusively to cross-networks, such as the spreading of misinformation from one social media to another and safety alerts found in downstream stages of the food supply chain. The complexity of cross-network interactions is further illustrated by an incident involving a malicious GitHub repository, as detailed in the upper part of Figure~\ref{fig: intro} and identified in \cite{GrowingBugsICSE21}, which was linked to over $40$ discussion threads on Stack Overflow. Questioners and less experienced users may be directed into using the alleged solution without maintaining a healthy skepticism. Once using the code from the malicious GitHub repository, the victims' devices might be compromised (e.g., system operations being disrupted). The challenge of tracing the origins and pathways of such misinformation is exacerbated in cross-network environments, where the initial propagation occurs in a network different from where the observations are made, as in the transition from GitHub to Stack Overflow. Furthermore, this complexity is compounded by multiple rounds of propagation and the possibility that contributors to Stack Overflow discussions may not intentionally disseminate misinformation, highlighting the need for more powerful source localization methodologies that account for cross-networks.
    
    
    Cross-network source localization is defined as locating diffusion sources from the source network by only giving the diffused observation of the target network, which is still under-explored. The challenge primarily lies in the separation between the networks where the diffusion originates and where it is observed, making traditional source localization techniques less effective due to the following critical obstacles. \textit{1) The difficulty of characterizing the distribution of diffusion sources only given the diffused observation of another network}. Understanding the distribution of potential sources is crucial for understanding the nature of diffusion processes and for quantifying the inherent uncertainties associated with identifying these sources \cite{ling2022source,ling2023deep,chowdhury4663083deep}. However, accurately learning the distribution of diffusion sources in a cross-network context requires the formulation of a conditional probability model that accounts for the observed diffusion within one network, given the structural and dynamical properties of another. This task requires fully considering different Network topologies, diverse node features, and varied propagation patterns, which makes the learning objective hard to model and optimize. \textit{2) The difficulty of jointly capturing dynamic and static features of the nodes in the cross-network.} Characterizing the distribution of diffusion sources is often conditioned on the intrinsic nature of the nodes and their connections. Existing works hardly leverage node features (e.g., text description and statistical node features) since entangling both types of features would lead to a high-dimensional and often intractable distribution of diffusion sources. Moreover, the nodes from different networks may have different intrinsic characteristics that help profile their diffusion dynamics and can predominantly help locate the sources.  \textit{3) The difficulty of jointly capturing the heterogeneous diffusion patterns of the cross-network.} Besides the difficulty of learning the distribution of diffusion sources, the diffusion patterns in both networks are also unknown to us. By correctly modeling the overall diffusion process, it is also essential to jointly consider the different propagation patterns of both networks. Additionally, the communication between different networks (i.e., cross-network propagation paths as noted in Figure~\ref{fig: intro}) also cannot be ignored.

    In this work, we propose the Cross-Network Source Localization (CNSL) method for locating the diffusion sources from a source network given its diffused observation from another target network under arbitrary diffusion patterns. Specifically, for the first challenge, we design a novel framework to approximate the distribution of diffusion sources by mean-field variational inference. For the second challenge, we propose a disentangled generative prior to encoding both static and dynamic features of nodes. For the last challenge, we model the unique diffusion dynamics of each network separately and integrate the learning process of these information diffusion models with the approximation of diffusion source distribution. This ensures accurate reconstruction of diffusion sources considering the specific propagation mechanisms of each network. We summarize our major contributions of this work as follows:
    \begin{itemize}[leftmargin=*]
        \item \textbf{Problem}. We design a novel formulation of the cross-network source localization and propose to leverage deep generative models to characterize the prior and approximate the distribution of diffusion sources via variational inference.
        \item \textbf{Technique}. We propose a unified framework to jointly capture 1) both static and dynamic node features, and 2) the heterogeneous diffusion patterns of both networks. The approximation of diffusion sources is fully aware of various node features and the interplay of cross-network information diffusion patterns.
        \item \textbf{Data}. Cross-network source localization lacks high-quality data, which is highly difficult to craft. We collect and curate a real-world dataset that accounts for the Cross-platform Communication Network, which records the real-world misinformation propagation from Github to Stack Overflow. We also provide a simulated cross-network dataset using agent-based simulation to disseminate misinformation across physical and social networks.
        \item \textbf{Experiments}. We conduct experiments against state-of-the-art methods designed originally for single-network source localization. Results show substantially improved performance of our method for cross-network source localization.
    \end{itemize}

    \section{Related Works}
    \noindent\textbf{Information Source Localization.}
    Diffusion source localization is defined as inferring the initial diffusion sources given the current diffused observation, which has attracted many applications, ranging from identifying rumor sources in social networks~\cite{jiang2016identifying} to finding blackout origins in smart grids \cite{shelke2019source}. Early approaches~\cite{prakash2012spotting, zhu2016information, zhu2017catch,wang2017multiple} focused on identifying the single/multiple source of an online disease under the Susceptible-Infected (SI) or Susceptible-Infected-Recover (SIR) diffusion patterns with either full or partial observation. Later on, Dong et al. \cite{dong2019multiple} further leverage GNN to enhance the prediction accuracy of LPSI. However, existing diffusion source localization methods cannot well quantify the uncertainty between different diffusion source candidates, and they usually require searching over the high-dimensional graph topology and node attributes to detect the sources, both drawbacks limit their effectiveness and efficiency. Moreover, in the past few years, more methods \cite{wang2022invertible,ling2022source,wang2023lightweight,qiu2023reconstructing,xu2024pgsl} have been proposed to address the dependency of prescribed diffusion models and characterize the latent distribution of diffusion sources, which have achieved state-of-the-art results. However, their methods still may not generalize to cross-network source localization due to the unique interconnected structure.
    
    \noindent\textbf{Information Diffusion on Cross Network.}
    The interconnection between cross-networks allows information to flow seamlessly from one platform to another through overlapping nodes. However, it is important to note that the patterns of influence and information propagation differ between various networks and can even vary within the same network. Recent studies in information diffusion across interconnected networks have made notable advancements. Earlier works \cite{khamfroush2016propagation, xuan2019self,deng2018modeling,ling2020nestpp} have developed different frameworks for correct modeling of the information flow within different network formats, such as wireless networks, social networks, and supply chains. Later on, many works have been proposed to study different features and applications of cross-networks, e.g., mitigating cascading failures \cite{tootaghaj2018mitigation, ghasemi2023robustness}. However, until today, there are few works \cite{do2024mim,ling2021forecasting} trying to correctly model the information diffusion pattern in the interconnected network system.

    
    \section{Cross-network Diffusion Source Localization}\label{sec: model}
    In this section, the problem formulation is first provided before deriving the overall objective from the perspective of the divergence-based variational inference. A novel optimization algorithm is then proposed to infer the seed nodes given the observed cross-network diffused pattern.
    
    \subsection{Problem Formulation}\label{sec: formulation}
    Cross-network $\mathcal{G} = (G_s, G_t)$ consists of a \textit{Source Network} $G_s=(V_s, E_s)$ and a \textit{Target Network} $G_t=(V_t, E_t)$. Both $G_s$ and $G_t$ are composed of a set of vertices $V_s$ and $V_t$ corresponding to individual users of the network as well as a set of edges $E_s\subseteq V_s\times V_s$ and $E_t\subseteq V_t\times V_t$ denote connecting pairs of users in both networks, respectively. In addition, $f_s\in \mathbb{R}^{N_s\times F}$ and $f_t\in \mathbb{R}^{N_t\times F}$ and denote the static features of both networks (e.g., associated text embedding, user age, social relations, etc.), where $F$ denotes the dimension of the node feature, and $N_t$, $N_s$ denote the number of nodes in each network, respectively. To connect the cross-network $\mathcal{G}$, there exists a set of bridge links between $G_s$ and $G_t$ denoted by $L = \{(v_s, v_t) | v_s \in V_s, v_t \in V_t\}$, which represent the propagation paths from $G_s$ to $G_t$.

    The information propagation in the cross-network is a \textit{one-directional} message passing from $G_s$ to $G_t$. More specifically, the propagation initiates from a group of nodes denoted as $x_s \in \{0, 1\}^{N_s}$ in the source network $G_s$, where each entry has a binary value representing whether the node is seed or not. After a certain period, the information propagates from $G_s$ to $G_t$ and infects a portion of nodes in $G_t$ through the bridge links $L$. We use $y_t\in[0, 1]^{N_t}$ to denote the infection probability of each node in $G_t$.
        
    \vspace{-0.1cm}
    \begin{problem}[\textbf{Cross-network diffusion source localization}]
    Given $\mathcal{G}$ and the diffused observation of the target network $y_t$, the problem of diffusion source localization in cross-networks (i.e., the inverse problem of diffusion estimation) requires finding $\Tilde x_s \in \{0, 1\}^{N_s}$ from the source network $G_s$, such that the empirical loss $||\Tilde x_s - x_s||_2^2$ is minimized, under the constraint that the diffused observation in the target graph $y_t$ could be generated from $\Tilde x_s$ through $L$.
    \end{problem}  \vspace{-0.1cm}

    However, reconstructing $\Tilde x_s$ from $y_t$ is difficult due to the following challenges. \textit{1) The difficulty of characterizing the distribution of seed nodes in the cross-network scenario}. To consider all possibilities of the seed nodes in cross-network source localization, it's desired to model the distribution of seed nodes $p(x_s)$ by characterizing the conditional probability $p(x_s|y_t)$. However, learning $p(x_s|y_t)$ requires jointly considering the topology structure of the cross-network $\mathcal{G}$ as well as the stochastic diffusion pattern through bridge links $L$. Existing works cannot be directly adapted due to the incapability of considering the complex cross-network scenario. \textit{2) The difficulty of jointly capturing dynamic and static features of the nodes in the cross-network}. The intrinsic patterns of the seed nodes consist of both dynamic patterns (i.e., the choice of seed nodes $x_s$) and static patterns (e.g., node features $f_s$). The correlated factors lead to the high-dimensional and often intractable distribution $p(x_s)$, which makes maximizing the joint likelihood $p(x_s, y_t)$ to be hard and computationally inefficient. \textit{3) The difficulty of jointly capturing the heterogeneous diffusion patterns of the cross-network}. The underlying diffusion process from $x_s$ to $y_t$ is not only affected by numerous factors (e.g., the infectiousness of the misinformation and the immunity power of the user), but the propagation patterns in the cross-network are inherently different in different networks.  
    
    \subsection{Latent Distribution Learning of Seed Nodes}
    To cope with the first challenge of characterizing the distribution of diffusion sources in the cross-network, we propose to utilize graph topology as well as the diffused observation to define the conditional probability $p(x_s|y_t)$. Since the diffused observation $y_t$ is conditioned on both networks $\mathcal{G}$ as well as the diffusion source $x_s$, we derive a conditional probability $p(y_t|x_s, \mathcal{G})\cdot p(x_s)$, where $p(x_s)$ is  the distribution of infection sources within $G_s$. To estimate the optimal diffusion source $\Tilde x_s$, we employ the Maximum A Posteriori (MAP) approximation by maximizing the following probability:
      \vspace{-0.1cm}
    \begin{align}
        \Tilde x_s = \argmax_{x_s} \:p(y_t|x_s, \mathcal{G})\cdot p(x_s) = \argmax_{x_s} \:p(x_s, y_t| \mathcal{G}). \nonumber
    \end{align}  \vspace{-0.0cm}
    However, since $p(x_s)$ is often intractable and entangles both static and dynamic features, we instead leverage deep generative models to characterize the implicit distribution of $p(x_s)$.

    To tackle the second challenge of jointly considering all static and dynamic node features, we propose a disentangled generative model to map the intractable and potentially high-dimensional $p(x_s)$ to latent embeddings in low-dimensional latent space. Specifically, we aim to learn the conditional distribution $p(x_s, y_t, \mathcal{G}|z_s,z_{fs})$ of $x_s$ given two latent variables $z_s$ and $z_{fs}$. Specifically, $z_s \in \mathbb{R}^{k_1}$ ($k_1 \ll N_s$) and $z_{fs} \in \mathbb{R}^{k_2}$ ($k_2 \ll N_s$) are obtained by an approximate posterior $p(z_s, z_{fs}|x_s, y_t, \mathcal{G})$, where $p(z_s, z_{fs})$ is the prior distribution of node's dynamic and static features. Note that $k_1$ and $k_2$ are the numbers of variables in each group, in order to capture the different types of semantic factors. 
    
\begin{figure*}
        \centering
        \includegraphics[width=0.9\textwidth]{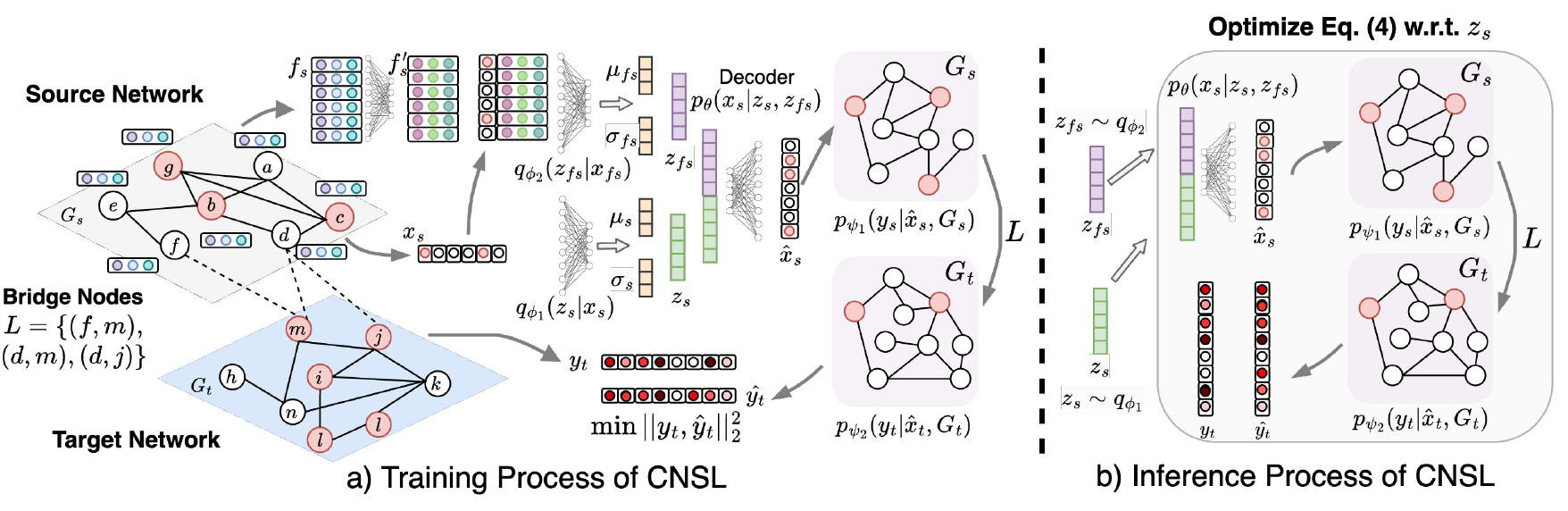}
        \vspace{-5mm}
        \caption{The training pipeline of CNSL  contains three steps: 1) $q_{\phi_1}$ and $q_{\phi_2}$ approximate the distribution of $p(z_s,z_{fs})$ in a disentangled manner; 2) the inferred latent variables $z_s$ and $z_{fs}$ are concatenated to reconstruct $\hat x_s$; 3) the reconstructed $\hat x_s$ is leveraged as initial seed nodes to initiate the cross-network information propagation and predict expected diffusion $\hat y_t$.}
        \label{fig:model_train}
        \vspace{-3.7mm}
    \end{figure*}

    The goal here is to learn the conditional distribution of $p(x_s)$ given $Z = (z_s, z_{fs})$, namely, to maximize the marginal likelihood of the observed cross-network diffusion in expectation over the distribution of the latent variable set $Z$ as $\mathbb{E}_{p_{\theta}(Z)}\left[p_{\theta}(x_s,y_t,\mathcal{G}|Z)\right]$. For a given observation of the information diffusion in the cross-network, the prior distribution of the latent representations $p(Z)$ is still intractable to infer. We propose solving it based on variational inference, where the posterior needs to be approximated by the distribution $q_{\phi}(Z|x_s, y_t, f_s, \mathcal{G})$. In this way, the goal becomes to minimize the Kullback–Leibler (KL) divergence between the true prior and the approximate posterior. Moreover, we assume $z_s$ and $z_{fs}$ capture different semantic factors. Specifically, $z_s$ is required to capture just the independent dynamic semantic factors of which nodes are infection sources, and $z_fs$ is required to capture the correlated semantic factors considering both dynamic features and static node features. To encourage this disentangling property of both posteriors, we introduce a constraint by trying to match the inferred posterior configurations of the latent factors to the prior $p(z_s,z_{fs})$ by setting each prior to being an isotropic unit Gaussian $\mathcal{N}(0,1)$, leading to the constrained optimization problem as:
    \begin{align}
    \max_{\theta,\phi}\quad&\mathbb{E}_{q_{\phi}(z_s,z_{fs}|x_s, y_t, \mathcal{G})}\left[p_{\theta}\left(x_s,y_t,\mathcal{G}|z_s,z_{fs}\right)\right],\nonumber \\ &\text{s.t.} \quad KL\left[q_{\phi}(z_s,z_{fs}|x_s, f_s, y_t, \mathcal{G})\vert\vert p(z_s,z_{fs})\right]<I \nonumber.
    \end{align}
    Furthermore, given the assumption that $p(z_s)$ represents the distribution of dynamic node features and $p(z_{fs})$ denotes the distribution of joint node features (entangles with both static and dynamic features), the constraint term can be decomposed as:
    \begin{align}
        &q_{\phi}(z_s,z_{fs}|x_s, f_s, y_t, \mathcal{G}) = q_{\phi_1}(z_s|x_s, y_t, \mathcal{G})\cdot q_{\phi_2}(z_{fs}|y_t, f_s, x_s, \mathcal{G})\nonumber
    \end{align}
    Then the objective function can be written as:
    
    \setlength{\abovedisplayskip}{-8pt} \setlength{\belowdisplayskip}{5pt}
    \begin{align}\label{eq:kl}
    \max_{\theta,\phi}\quad&\mathbb{E}_{q_{\phi}(z_s,z_{fs}|x_s, y_t, \mathcal{G})}\left[p_{\theta}\left(x_s,y_t,\mathcal{G}|z_s,z_{fs}\right)\right], \\ &\text{s.t.} \quad KL\left[q_{\phi_1}(z_s|x_s, y_t, \mathcal{G})\vert\vert p(z_s)\right]<I_s, \nonumber\\
    &\quad \quad KL\left[q_{\phi_2}(z_{fs}|y_t, x_s, \mathcal{G})\vert\vert p(z_{fs})\right]<I_{fs},\nonumber
    \end{align}
    where we decompose $I$ into two separate parts (i.e., $I_s$ and $I_{fs}$) of the information capacity to control each group of latent variables so that the variables inside each group of latent variables are disentangled. In practice, $q_{\phi_1}(\cdot)$ and $q_{\phi_2}(\cdot)$ are implemented as two encoders with multi-layer perceptron structure. More details can be found in Figure~\ref{fig:model_train}. 
    \setlength{\abovedisplayskip}{3pt} \setlength{\belowdisplayskip}{3pt}

    \subsection{Cross Network Diffusion Model Learning}
    To address the third challenge, i.e., making the source localization be aware of the heterogeneous diffusion patterns between networks, locating diffusion origins $x_s$ may not only involve estimating the distribution of seed nodes but the process should also be determined by correctly modeling the information diffusion across diverse and interlinked network structures $\mathcal{G}$. In the context of cross-network information diffusion, the diffused observation $y_t$ is determined by the diffusion source $x_s$ under the cross-network $\mathcal{G}$ through bridge links $L$. Therefore, the conditional distribution $p_{\theta}(x_s,y_t, x_{f},\mathcal{G}|z_s,z_{fs})$ can further be decoupled as:
    \begin{equation*}
        \log p_{\theta}(x_s,y_t, x_{f},\mathcal{G}|z_s,z_{fs}) = \log [p_{\psi}(y_t|x_s, \mathcal{G})] + \log[p_{\theta}(x_s|z_s,z_{fs})],
    \end{equation*}
    where $p_{\psi}(\cdot)$ models the probability of the infection status $y_t$ of nodes in $G_t$ given seed nodes $x_s$ in $G_s$. Moreover, the second term $p_{\theta}(x_s|z_s,z_{fs})$  reveals that the latent variables $Z$ only encodes information from $x$ (i.e., $y_t \bot Z|x_s, \mathcal{G}$). According to the assumption, we could also simplify both encoders as $q_{\phi_1}(z_s|x_s, \mathcal{G})$ and $q_{\phi_2}(z_{fs}|x_s, f_s,\mathcal{G})$ in Eq. \eqref{eq:kl} by removing $y_t$ from the input.\\

    \noindent\textbf{Cross-network Information Propagation.} Modeling the diffusion from $x_s$ to $y_t$ is complex due to multiple influencing factors, such as misinformation's infectiousness and the distinct propagation patterns across networks like GitHub and Stack Overflow, which cater to different user communities. The unknown nature of these diffusion patterns prevents the use of standard models like Linear Threshold or Independent Cascade. This complexity underlines the need to decompose and simplify $p_{\psi}(y_t|x_s, \mathcal{G})$ to analyze the diverse diffusion behaviors in $G_s$ and $G_t$ through a learning approach:
    \begin{align}\label{eq:decouple}
        \log p_{\psi}(y_t|x_s, \mathcal{G}) = \log p_{\psi_1}(y_s|x_s, G_s) + \log p_{\psi_2}(y_t|y_s, x_t, G_t).
    \end{align}
    In this simplified decomposition, $p_{\psi_1}(\cdot)$ characterizes the diffusion pattern of $G_s$ given the seed nodes $x_s$, which is independent of the information propagation in $G_t$. $y_s \in [0,1]^{N_s}$ records the infection status of all nodes in the source network $G_s$. When the diffusion is complete in $G_s$, the infection probability is directly transferred to the target network $G_t$ through bridge links $L$ so that some nodes in $G_t$ have initial infection status (denoted as $x_t$) to initiate the infection process in $G_t$. The propagation in $G_t$ is then modeled by $p_{\psi_2}(y_t|y_s, x_t, G_t)$ by taking the graph structure $G_t$ and initial seed infection probability $x_t$ as inputs. More details of the derivation is provided in the Appendix.

    \noindent\textbf{Monotonic Constraint on Information Diffusion.} The information diffusion on the regular network is often regularized by the monotone increasing property~\cite{dhamal2016information,ling2022source}. In this work, we also assume the same monotonic property holds in the cross-network information diffusion, namely $y_t^{(i)} \succeq y_t^{(j)},\: \forall \: x_s^{(i)} \supseteq x_s^{(j)}$. Specifically, selecting more seed nodes in $G_s$ would result in a generally higher (or at least equal) infection probability of nodes in $G_s$ according to the property of diminishing returns. Subsequently, the bridge links would transfer the infection probability from $y_s$ to $x_t$, and similarly, the probability of each node being infected in $y_t^{(i)}$ (estimated from $x_t^{(i)}$) should be greater or equal to $y_t^{(j)}$ (estimated from $x_t^{(j)}$), such that $y_t^{(i)} \succeq y_t^{(j)}$. Therefore, owing to the monotonic increasing property of the information diffusion, we add the constraint $\lambda \big|\big|\max(0, y_t^{(j)}-y_t^{(i)})\big|\big|_2^2, \forall \: x_s^{(i)} \supseteq x_s^{(j)},$ to Eq. \eqref{eq:kl},
    where we transform the inequality constraint into its augmented Lagrangian form to minimize $\rVert\max(0, y_t^{(j)}-y_t^{(i)})\big|\big|_2^2$ and $\lambda > 0$ denotes regularization hyperparameter.\\
    \noindent\textbf{Overall Objective for Training.} The training procedure of the proposed CNSL model is coupled with Eq. \eqref{eq:kl}, Eq. \eqref{eq:decouple}, and the monotonic increasing constraint: 
    \begin{align}\label{eq:objective}\vspace{-3mm}
\mathcal{L}_{\text{train}}&=\max_{\theta,\phi_1,\phi_2}\:\mathbb{E}_{q_{\phi}}\left[p_{\theta}(x_s,y_t, x_{f},\mathcal{G}|z_s,z_{fs})\right], \\ &\quad\text{s.t.} \quad KL\left[q_{\phi_1}(z_s|x_s, \mathcal{G})\vert\vert p(z_s)\right]<I_s, \nonumber\\
    &\quad\quad\quad KL\left[q_{\phi_2}(z_{fs}|x_s, f_s, \mathcal{G})\vert\vert p(z_{fs})\right]<I_{fs},\nonumber\\
    &\quad\quad\quad y_t^{(i)} \succeq y_t^{(j)},\: \forall \: x_s^{(i)} \supseteq x_s^{(j)},\nonumber\\    &\nonumber=\min_{\theta,\phi_1,\phi_2,\psi_1,\psi_2}\:-\mathbb{E}_{q_{\phi}}\big[\log p_{\theta}(x_s|z_s,z_{fs})\\&\nonumber\quad\quad+\log p_{\psi_1}(y_s|x_s, G_s)+\log p_{\psi_2}(y_t|y_s, x_t, G_t)\big],\\
    &\quad\text{s.t.} \quad KL\left[q_{\phi_1}(z_s|x_s, \mathcal{G})\vert\vert p(z_s)\right]<I_s, \nonumber\\
    &\quad\quad\quad KL\left[q_{\phi_2}(z_{fs}|x_s, f_s, \mathcal{G})\vert\vert p(z_{fs})\right]<I_{fs},\nonumber\\
    &\quad\quad\quad \big|\big|\max(0, y_t^{(j)}-y_t^{(i)})\big|\big|_2^2\nonumber, 
    \end{align}  
    where we only need to sample one $x_s^{(i)}$ and many $x_s^{(j)}$'s (such that $x_s^{(i)} \supseteq x_s^{(j)}$) as training samples for each mini-batch. The $y_t^{(i)}$ and $y_t^{(j)}$'s are estimated by arbitrary diffusion patterns. For simplicity, we omit the subscript of $\mathbb{E}_{q_{\phi}(z_s,z_{fs}|x_s, y_t, \mathcal{G})}$ as $\mathbb{E}_{q_{\phi}}$ when the context is clear. The overall framework is summarized in Figure~\ref{fig:model_train}.
    \vspace{-0.12cm}
    \subsection{Cross-network Seed Set Inference}    \vspace{-0.1cm}
    Upon training completion, the joint probability $p(z_s, z_{fs})$ is approximated by the posterior $q_{\phi}(z_s,z_{fs}|x_s, f_s, y_t, \mathcal{G})$. Both $p_{\psi_1}(\cdot)$ and $p_{\psi_2}(\cdot)$ effectively classify the diffusion patterns across networks. This study introduces a sampling method for $\Tilde x_s \sim p(x_s)$ by marginalizing over $p(z_s) \cdot p(z_{fs})$ to conduct MAP estimation, where $p(x_s)=\sum_{z_s}\sum_{z_{fs}}p_{\theta}(x_s|z_s, z_{fs}) p(z_s, z_{fs})$. However, marginalizing the standard Gaussian prior $p(z_s, z_{fs})$ necessitates extensive sampling to align the sample distribution with the target distribution, increasing computational load. Additionally, it is also hard to sample individual latent variables from the joint distribution of $p(z_s, z_{fs})$. To cope with both challenges, we consider the density over the inferred latent variables induced by the approximate posterior inference mechanism, and we propose the following objective w.r.t. $z_s$ to infer $\Tilde x_s$ in an optimized manner.
    Specifically, the inference objective function $\mathcal{L}_{\text{pred}}$ is written as:
    \begin{align}\label{eq: infer}
        \mathcal{L}_{\text{pred}} &= \max_{z_s} \:\mathbb{E}\left[p_{\psi}(y_t|x_s, \mathcal{G})\cdot p_{\theta}(x_s|z_s, z_{fs})\right],\\
        &\quad\text{s.t.}\quad z_s \sim q_{\phi_1}(z_s|\hat x_s, \mathcal{G}), \quad z_{fs} \sim q_{\phi_2}(z_{fs}|\hat x_s, f_s, \mathcal{G})\nonumber,\\
        &=\min_{z_s}\:-\mathbb{E}\Big[\log p_{\psi}(y_t|x_s, \mathcal{G})+\log \big[\sum\nolimits_{z_s}\sum\nolimits_{\hat x_s}p_{\theta}(x_s|z_s, z_{fs})\big] \Big]\nonumber\\
        &\quad\text{s.t.}\quad z_s \sim q_{\phi_1}(z_s|\hat x_s, \mathcal{G}), \quad z_{fs} \sim q_{\phi_2}(z_{fs}|\hat x_s, f_s, \mathcal{G})\nonumber,\vspace{-5mm}
    \end{align}
    where we sample many $\hat x_s$ from the training set, and obtain equal amount of $z_s$ from $q_{\phi_1}(\cdot)$. Note that we optimize $z_s$ (dynamic latent variable) only, instead of both $z_s$ and $z_{fs}$ (static-dynamic entangled latent variable), which is rooted in the specific roles these variables play in the model. $z_s$ is targeted for optimization because it encodes dynamic information crucial for identifying better seed nodes in the context of information diffusion. This dynamic aspect is mutable and can be optimized to improve source localization accuracy. On the other hand, $z_{fs}$ entangles both dynamic and static information, where the static part represents unchangeable node features. Optimizing $z_{fs}$ would be less efficient because static features, by their nature, cannot be optimized. The optimization process aims to adjust variables to improve model performance, but since static features remain constant, attempting to optimize $z_{fs}$ would not enhance the model's ability to localize diffusion sources. 

    \noindent\textbf{Implementation of the Seed Set Inference.} We provide implementation details of the overall inference process here. Specifically, the inference framework first samples $k$ seed node set $\hat x_s$ from the training set, and we can take the average value $\bar z_s$ and $\bar z_{fs}$ from the learned latent distributions with taking $k$ different $\hat x_s$ as input:
    \begin{align}\label{eq: z_sample}
        \bar z_s &= \frac{1}{k} \sum\nolimits^k_i q_{\phi_1}( z_s|\hat x^{(i)}_s, \mathcal{G}),
        \bar z_{fs} = \frac{1}{k} \sum\nolimits^k_i q_{\phi_2}(z_s|\hat x^{(i)}_s, f_s, \mathcal{G}).
    \end{align}
    We concatenate $\bar z_s$ and $\bar z_{fs}$ as input to minimize the inference loss in Eq. \ref{eq: infer}. The latent variable $z_s$ is iteratively optimized according to the inference objective function to minimize $-\log p_{\psi}(y_t|x_s, \mathcal{G})$. In practice, Eq. \eqref{eq: infer} cannot be optimized directly, we thus provide a practical version of the inference objective function: since the diffused observation $y_t$ fits the Gaussian distribution and the seed set $x_s$ fits the Bernoulli distribution, we can simplify Eq. \eqref{eq: infer} as:
    \begin{align}\label{eq: infer_2}
        \mathcal{L}_{\text{pred}} = \min_{z_s} &-\Big[\log \big[\prod\nolimits_{i=0}^{N_s}f_{\theta}(z_s^{(i)}, z_{fs}^{(i)})^{x_s^{(i)}}(1-f_{\theta}(z_s^{(i)}, z_{fs}^{(i)})^{1-x_s^{(i)}}\big]\nonumber\\
        &+\big\rVert \Tilde y_t- y_t\big\rVert_2^2 \Big]
    \end{align}
    where the $\Tilde y$ is given as the optimal influence spread (i.e., $\Tilde y_t = N_t$). In other words, the inference objective is guided by the discrepancy between the inferred $y_t$ and the ground truth $\Tilde y_t$. We visualize the overall inference procedure in Figure~\ref{fig:model_train} (b). Specifically, we sample $\bar z_{fs}$ and $\bar z_s$, according to Eq. \eqref{eq: z_sample}, and leverage $p_{\theta}(\cdot)$ to decode $\hat x_s$. The predicted $\hat x_s$ is leveraged to initiate the cross-network diffusion and predict $\hat y_t$. The optimization supervision consists of 1) the mean squared loss between $\hat y_t$ and the ground truth $y_t$ as well as 2) the probability of node $v_i$ being seed node $f_{\theta}(z^{(i)}_s, z^{(i)}_{fs})\in [0, 1]$.

\section{Experimental Evaluation}
This section reports both qualitative and quantitative experiments that are carried out to test the performance of CNSL and its extensions on a simulated dataset that simulates the spread of misinformation across a city-level population and a collected real-world cross-network dataset obtained by crawling two online networking platforms and cross-references between them.

\subsection{Real-world Dataset: Cross-Platform Communication Network}
We collected real-world data from GitHub and Stack Overflow to form the cross-platform communication network, where information flows from Github to Stack Overflow since many posts in Stack Overflow have mentioned or discussed Github Repositories when addressing user's questions. We started by downloading the Stack Overflow public data dump provided by the Internet Archive. Then, we extracted all the Stack Overflow posts where their post texts contain a URL to GitHub (i.e., 439,753 posts mapping to 439,753 repositories). We further built the Stack Overflow network by finding the question posts, answer posts, and related posts of the current 439,753 posts. This yielded a total of 1,410,600 Stack Overflow posts, encompassing data from 2008 up to 2023.

To obtain the GitHub network, we expanded our initial GitHub network by finding all GitHub repositories that the existing repositories depended upon. We utilized an open-source tool\footnote{\url{https://github.com/edsu/xkcd2347}}, which uses the GitHub GraphQL API to obtain the dependency information. The resulting GitHub network contains 533,240 repositories. For our experiment, we sampled GitHub repositories from the year 2021 and their dependent repositories from the year before 2021 (i.e., 1204 nodes and 1043 edges). We then found their corresponding Stack Overflow posts (i.e., 3862 nodes and 3149 edges). We obtained the ground truth in a pseudo-setting: we randomly sampled 10\% of the GitHub nodes as seed nodes, and simulated their diffusion process within the GitHub network and the Stack Overflow network (i.e., 120 GitHub seed nodes, 354 GitHub infected nodes, 195 Stack Overflow seed nodes, and 482 Stack Overflow infected nodes). 

\vspace{-0.1cm}
\subsection{Simulated Dataset: Agent-Based Geo-Social Information Spread} \label{subsec: g2s_data_gen}
 We leverage and agent-based simulation framework based on realistic Patterns of Life ~\cite{kim2020,kim2019simulating} to simulate the spread of misinformation across social and physical networks. In this simulation, an agent represents a simulated individual who commutes to their workplace, eats at restaurants, and meets friends and recreational sites. Inspired by the Theory of Planned Behavior~\cite{ Ajzen1991} and Maslow's Hierarchy of Needs~\cite{maslow1943theory} as theories of human behavior, agents are driven by physiological needs to eat and have a shelter, safety needs such as financial stability requiring them to go to work, and needs for love requiring them to meet friends and build and maintain a social network. Details of the theories of social science informing this simulation are found in~\cite{zufle2023urban} and details to use this simulation for data-generation are described in~\cite{amiri2023massive}. 
 
 We augmented this simulation framework to simulate the spread of misinformation using a simple Susceptible-Infectious disease model. The simulation is initialized with 15,000 agents. A small number of $n$ (by default, $n=5$) agents are selected randomly as the sources of misinformation and flagged as ``Infectious'' and all other agents are initially flagged as ``Susceptible''. Agents can spread misinformation in two ways: 1) through collocation, allowing an agent to spread the misinformation in-person to other agents located at the same workplace, restaurant, or recreational site, and 2) through the social network, allowing an agent to spread misinformation to their friends regardless of their location.
To allow the generation of large datasets for source localization, each spreading misinformation is stopped after five simulation days. At this time, the following datasets are recorded:
\begin{itemize}[leftmargin=*]
   \item \textit{Ground Truth.} The set of $n$ agents that were initially seeded with the misinformation.
   \item \textit{Misinformation Spread.} The list of agents to whom the misinformation has spread after five days.
    \item \textit{The Complete Co-location Network.} This network captures the agents who meet each other and thus, may spread misinformation through co-location.
    \item \textit{The Observed Co-location Network.} This network is a randomly sampled subset of agents from the complete co-location network. It represents the agents in the complete co-location network that are parts of the simulated location tracking. This network is used to simulate the realistic case of not having access to the location data of every individual.
    \item \textit{The Complete Social Network.} This network records the friend and family connections of all agents which may infect each other through social contagion. 
    \item \textit{The Observed Social Network.} This network includes a randomly sampled subset of agents from the complete social network and simulates the social media environment. This network simulates the realistic case where an observed social media network may not capture the entire population.
    \item \textit{Cross-Network Links through Identity.} Links between the two observed networks are defined through identity. Any individual agent in the co-location network is (trivially) connected to itself in the social network.
\end{itemize}

\begin{table*}[h]
    \centering
    \resizebox{0.8\textwidth}{!}{%
    \begin{tabular}{@{}llcccccccccccc@{}}
        \toprule
        & & \multicolumn{4}{c}{\textbf{LT2LT}} & \multicolumn{4}{c}{\textbf{LT2IC}} & \multicolumn{4}{c}{\textbf{LT2SIS}} \\
        \cmidrule(lr){3-6} \cmidrule(lr){7-10} \cmidrule(lr){11-14}
        \textbf{Category} & \textbf{Method} & \textbf{PR} & \textbf{RE} & \textbf{F1} & \textbf{AUC} & \textbf{PR} & \textbf{RE} & \textbf{F1} & \textbf{AUC} & \textbf{PR} & \textbf{RE} & \textbf{F1} & \textbf{AUC} \\
        \midrule
        \multirow{2}{*}{Rule-based} 
        & LPSI & 0.156 & 0.841 & 0.263 & 0.583 & 0.141 & 0.849 & 0.242 & 0.533 & 0.079 & 0.942 & 0.127 & 0.497 \\
        & OJC  & 0.104 & 0.035 & 0.052 & 0.500 & 0.116 & 0.036 & 0.054 & 0.502 & 0.113 & 0.036 & 0.053 & 0.501 \\\midrule
        \addlinespace
        \multirow{3}{*}{\begin{tabular}[c]{@{}l@{}}Learning\\ based\end{tabular}}
        & GCNSI & 0.103 & 0.858 & 0.184 & 0.636 & 0.103 & 0.866 & 0.184 & 0.622 & 0.114 & 0.801 & 0.199 & 0.635 \\
        & IVGD  & 0.228 & 0.948 & 0.368 & 0.139 & 0.227 & 0.874 & 0.359 & 0.138 & 0.123 & 0.985 & 0.215 & 0.240 \\
        & SL-VAE & 0.249 & 0.947 &	0.395 & 0.703 & 0.192 & 0.847 &	0.313 &	0.689 & 0.242 & 0.931 &	0.385 &	0.612 \\
        & DDMSL & 0.251 & 0.923 & 0.394 & 0.815 & 0.309 & 0.845 &	0.454 &	0.732 & 0.320 & 0.842 &	0.464 &	0.772 \\\midrule
        \addlinespace
        \multirow{2}{*}{Our Method} & CNSL & \textbf{0.332} & \textbf{0.996} & \textbf{0.498} & \textbf{0.888} 
        & \textbf{0.332} & \textbf{0.997} & \textbf{0.498} & \textbf{0.889} 
        & \textbf{0.332} & \textbf{0.997} & \textbf{0.498} & \textbf{0.890}\\
        & CNSL-W/O & 0.103 & 0.922 & 0.185 &  0.520 
        & 0.103 & 0.930 & 0.186 &  0.511
        & 0.103 & 0.917 & 0.186 & 0.517 \\
        \bottomrule
    \end{tabular}%
    }
        \caption{Performance comparison for cross-platform communication network under LT diffusion pattern for the first network with LT, IC, and SIS diffusion pattern for the second network.}
    \label{tab: enhanced_lt_comparison}
        \vspace{-6mm}
\end{table*}

\begin{table*}[h]
    \centering
    \resizebox{0.8\textwidth}{!}{%
    \begin{tabular}{@{}llcccccccccccc@{}}
        \toprule
        & & \multicolumn{4}{c}{\textbf{IC2LT}} & \multicolumn{4}{c}{\textbf{IC2IC}} & \multicolumn{4}{c}{\textbf{IC2SIS}} \\
        \cmidrule(lr){3-6} \cmidrule(lr){7-10} \cmidrule(lr){11-14}
        \textbf{Category} & \textbf{Method} & \textbf{PR} & \textbf{RE} & \textbf{F1} & \textbf{AUC} & \textbf{PR} & \textbf{RE} & \textbf{F1} & \textbf{AUC} & \textbf{PR} & \textbf{RE} & \textbf{F1} & \textbf{AUC} \\
        \midrule
        \multirow{2}{*}{Rule-based} 
        & LPSI & 0.124 & 0.868 & 0.217 & 0.489 & 0.215 & 0.657 & 0.324 & 0.562 & 0.129 & 0.906 & 0.226 & 0.522 \\
        & OJC  & 0.117 & 0.032 & 0.050 & 0.503 & 0.097 & 0.027 & 0.042 & 0.499 & 0.115 & 0.032 & 0.050 & 0.502\\\midrule
        \addlinespace
        \multirow{3}{*}{\begin{tabular}[c]{@{}l@{}}Learning\\ based\end{tabular}} 
        & GCNSI & 0.142 & 0.638 & 0.233 & 0.623 & 0.170 & 0.476 & 0.251 & 0.627 & 0.152 & 0.602 & 0.243 & 0.630 \\
        & IVGD  & 0.120 & 0.979 & 0.210 & 0.733 & \textbf{0.548} & 0.391 & 0.083 & 0.439 & 0.115 & 0.825 & 0.195 & 0.733 \\
        & SL-VAE & 0.254 & 0.881 &	0.394 &	0.719 & 0.195 & 0.909 &	0.321 &	0.703 & 0.185 & 0.829 & 0.302 &	0.592 \\
        & DDMSL & 0.286 & 0.827 & 0.425 & 0.818 & 0.318 & 0.886 &	0.468 &	0.753 & 0.270 & 0.833 & 0.408 &	0.689 \\\midrule
        \addlinespace
        \multirow{2}{*}{Our Method} & CNSL & \textbf{0.333} & \textbf{0.990} & \textbf{0.498} & \textbf{0.887} 
        & 0.333 & \textbf{0.998} & \textbf{0.499} & \textbf{0.891} 
        & \textbf{0.332} & \textbf{0.997} & \textbf{0.498} &  \textbf{0.888}\\
         & CNSL-W/O & 0.103 & 0.922 & 0.186 & 0.514
        & 0.103 & 0.935 & 0.185 &  0.515
        & 0.103 & 0.928 & 0.185 &  0.516\\
        \bottomrule
    \end{tabular}%
    }    
    \caption{Performance comparison for cross-platform communication network under IC diffusion pattern for first network with LT, IC, and SIS diffusion pattern for the second network.}    
    \label{tab: enhanced_ic_comparison}
    \vspace{-6mm}
\end{table*}

\begin{table*}[h]
    \centering
    \resizebox{\textwidth}{!}{%
    \begin{tabular}{@{}llcccccccccccccccc@{}}
        \toprule
        & & \multicolumn{4}{c}{\textbf{G2S-A-D0}} & \multicolumn{4}{c}{\textbf{G2S-B-D0}} & \multicolumn{4}{c}{\textbf{G2S-A-D1}} & \multicolumn{4}{c}{\textbf{G2S-B-D1}}\\
        \cmidrule(lr){3-6} \cmidrule(lr){7-10} \cmidrule(lr){11-14} \cmidrule(lr){15-18}
        \textbf{Category} & \textbf{Method} & \textbf{PR} & \textbf{RE} & \textbf{F1} & \textbf{AUC} & \textbf{PR} & \textbf{RE} & \textbf{F1} & \textbf{AUC} & \textbf{PR} & \textbf{RE} & \textbf{F1} & \textbf{AUC} & \textbf{PR} & \textbf{RE} & \textbf{F1} & \textbf{AUC} \\
        \midrule
        \multirow{2}{*}{Rule-based} 
        & LPSI & 0.147 & 0.982 & 0.256 & 0.512 & 0.165 & 0.954 & 0.281 & 0.609 & 0.152 & 0.903 & 0.260 & 0.475 & 0.224 & 0.973 & 0.364 & 0.578 \\
        & OJC  & 0.053 & 0.018 & 0.022 & 0.496 & 0.125 & 0.039 & 0.051 & 0.507 & 0.063 & 0.040 & 0.043 & 0.497 & 0.115 & 0.058 & 0.071 & 0.505 \\\midrule
        \addlinespace
        \multirow{3}{*}{\begin{tabular}[c]{@{}l@{}}Learning\\ based\end{tabular}}
        & GCNSI & 0.123 & \textbf{1.000} & 0.216 & 0.744 & 0.117 & \textbf{1.000} & 0.207 & 0.351 & 0.183 & \textbf{1.000} & 0.300 & 0.250 & 0.221 & \textbf{1.000} & 0.341 & 0.193 \\
        & IVGD  &  0.139 & \textbf{1.000} & 0.244 & 0.502 & 0.138 & \textbf{1.000} & 0.242 & 0.500 & 0.218 & \textbf{1.000} & 0.352 & 0.490 & 0.266 & \textbf{1.000} & 0.409 & 0.500 \\
        & SL-VAE & 0.364 & 0.863 & 0.512 & 0.707 & 0.289 & 0.788 & 0.423 & 0.611 & 0.289 & 0.754 & 0.418 & 0.664 & 0.425 & 0.893 & 0.576 & 0.725 \\\midrule
        \addlinespace
        \multirow{2}{*}{Our Method} & CNSL & \textbf{0.481} & 0.816 & \textbf{0.605} & \textbf{0.931} 
        & \textbf{0.452} & 0.885 & \textbf{0.598} & \textbf{0.933} & \textbf{0.499} & 0.779 & \textbf{0.609} & \textbf{0.894} 
        & \textbf{0.539} & 0.987 & \textbf{0.698} & \textbf{0.901}\\
         & CNSL-W/O S & 0.122 & 1.000 & 0.219 & 0.503 
        & 0.117 & 1.000 & 0.2101 & 0.488 & 0.183 & 0.998 & 0.309 & 0.499 
        & 0.221 & 0.999 & 0.362 & 0.501\\
        \bottomrule
    \end{tabular}%
    }    
    \caption{Performance comparison for Geo-Social Information Spread Data (G2S) for two types (A, B) of simulation. Here $D0$ considers the initial sources of misinformation as seed nodes and $D1$  considers the initial sources of misinformation and the infected agents at the first day as seed nodes.}    
    \label{tab: enhanced_sim_comparison}
    \vspace{-7mm}
\end{table*}

Once this data is collected, the misinformation spread status of all agents is set to ``Susceptible'' and $n$ new agents are selected as the seed nodes of a new case of misinformation. This process of creating new cases of misinformation is iterated every five simulation days to create an unlimited number of realistic datasets of information spread across the physical and social spaces.

For the dataset used for the following experiments, there are 5,281 agents and 8,276 edges in the observed co-location network, and 5,669 agents and 17,948 edges in the observed social network. Each case of misinformation spread yields between 50-200 agents to which the misinformation spreads after five days. 
This synthetic dataset allows us to capture realistic misinformation spread across both networks. Due to some agents not being captured in the two networks, this dataset allows us to simulate the realistic case where misinformation may spread outside of the observed networks. We provide the code for our agent-based misinformation simulation framework in a GitHub repository found at \url{https://github.com/Siruiruirui/misinformation}. This repository also contains the generated dataset used in the following experiments.


\subsection{Experiment Setup}
\noindent\textbf{Implementation Details.} We employ a two-layer MLP for learning node features, which are concatenated with the seed vector in the subsequent stage before being input to the encoder $q_{\phi_{2}}(\cdot)$. Both encoders ($q_{\phi_{1}}(\cdot)$, $q_{\phi_{2}}(\cdot)$) and the decoder $p_{\theta}(\cdot)$ utilize three-layer MLPs with non-linear transformations. We use GNN model architecture coupled with a two-layer MLP network as the aggregation network with $64$ hidden units for the two propagation models ($p_{\psi_1}(\cdot)$ and $p_{\psi_2}(\cdot)$). The learning rates for encoder-decoder, $p_{\psi_1}(\cdot)$, and $p_{\psi_2}(\cdot)$ are set to 0.0001, 0.005, and 0.01 respectively in a multi-optimization manner. Additionally, the number of epochs is $15$ for all datasets, with a batch size of $2$. The iteration numbers for inference are set to $2$ for all datasets.

\noindent\textbf{Comparison Methods.} We illustrate the performance of CNSL in various experiments against two sets of methods: 1) \textbf{Rule-based methods}: \textit{LPSI}~\cite{wang2017multiple} predicts the rumor sources based on the convergent node labels without the requirement of knowing the underlying information propagation model; \textit{OJC}~\cite{zhu2017catch} aims at locating sources in networks with partial observations, which has strength in detecting network sources under the SIR diffusion pattern. 2) \textbf{Learning-based methods}: \textit{GCNSI} \cite{dong2019multiple} learns latent node embedding with GCN to identify multiple rumor sources close to the actual source; \textit{IVGD} \cite{wang2022invertible} propose a graph residual model to make existing graph diffusion models invertible; \textit{SL-VAE} \cite{ling2022source} proposed to learn the graph diffusion model with a generative model to characterize the distribution of diffusion sources. DDMSL \cite{yan2024diffusion} proposed a diffusion model-based source localization method to recover each diffusion step iteratively. Note that existing comparison methods are not designed for cross-network source localization, in order to conduct a fair comparison, we repeated each model separately for two networks and learned the two networks. We used bridge links $L$ to connect these two models.

\noindent\textbf{Evaluation Metrics.} Source localization is a classification task so that we use two main metrics to evaluate the performance of our proposed model: \emph{1). F1-Score (F1)} and \emph{2). ROC-AUC Curve (AUC)}, as they are classical metrics for classification tasks. since most real-world scenarios tend to have an imbalance between the number of diffusion sources and non-source nodes (fewer diffusion sources), we additionally leverage PR@100 to evaluate the precision of the top-100 prediction returned by models.


\subsection{Quantitative Analysis} \label{subsec:qa}
We evaluated the models in different diffusion configurations. For the cross-platform communication data, the underlying diffusions are LT (Table \ref{tab: enhanced_lt_comparison}) and IC (Table \ref{tab: enhanced_ic_comparison}) for the first network which was followed by other three diffusion patterns (LT, IC, and SIS) for the second network in each case. For the Geo-Social information spread data (Table \ref{tab: enhanced_sim_comparison}), the underlying diffusion pattern has been explained in Section \ref{subsec: g2s_data_gen}. For that dataset, we used two different simulations (A and B) and also used two different types of seed selections. Here $D0$ considers the initial sources of misinformation as seed nodes and $D1$  considers the initial sources of misinformation and the infected agents on the first day as seed nodes.

\noindent\textbf{Performance in the cross-platform communication network.} 
Table \ref{tab: enhanced_lt_comparison} shows that CNSL excels others across all metrics and diffusion patterns. In the first network with LT diffusion pattern (LT2LT, LT2IC, LT2SIS), CNSL achieves the highest recall (RE) in all scenarios, with scores of 0.996, 0.997, and 0.997, respectively, indicating its superior ability to identify all relevant instances in the dataset. Additionally, CNSL also exhibits the best precision (PR) in LT2LT and LT2IC scenarios, and competitive precision in the LT2SIS scenario. The F1 scores, which balance precision and recall, are also highest for CNSL, peaking at 0.498 in both LT2LT and LT2IC patterns, demonstrating the method's overall efficiency and accuracy. The AUC scores for CNSL are robust, ranking highest in LT2LT and LT2SIS scenarios, signifying excellent model performance across various threshold settings.
In the Table \ref{tab: enhanced_ic_comparison}  first network with IC diffusion pattern (IC2LT, IC2IC, IC2SIS), CNSL's performance remains impressive, maintaining the highest recall scores of 0.990, 0.998, and 0.997, respectively. CNSL also boasts the highest F1 scores in all scenarios, with a notable 0.499 in IC2IC, suggesting a balanced performance between precision and recall. The AUC scores for CNSL are again the highest, with 0.887 in IC2LT and 0.891 in IC2IC, indicating its strong discriminative ability. Overall, CNSL demonstrates considerable strength in reliably identifying relevant instances across different diffusion patterns and networks, while maintaining high precision and excellent area under the ROC curve.

\noindent\textbf{Performance in geo-social information spread data.}
In Table \ref{tab: enhanced_sim_comparison}, the performance of various methods on Geo-Social Information Spread Data (G2S) is evaluated for two simulation types, A and B, with two different seeding strategies, D0 and D1. Our method, CNSL, exhibits strong performance across all scenarios. In the G2S-A-D0 simulation, CNSL achieves a high precision (PR) of 0.481, showing its effectiveness in correctly identifying misinformation spread. It also has the highest F1 score of 0.605 and an AUC of 0.931, indicating a balanced precision-recall trade-off and excellent model discrimination ability, respectively.
For the G2S-B-D0 simulation, CNSL's precision (0.452) and F1 score (0.598) are notable, and the AUC of 0.933 is the highest compared to other methods, suggesting CNSL's consistency and reliability. In the G2S-A-D1 scenario, CNSL maintains a high recall (RE) of 0.779 and an impressive AUC of 0.894, which signifies its capacity to identify true misinformation cases effectively when the seeding includes infected agents from the first day.
Remarkably, in the G2S-B-D1 scenario, CNSL stands out with the highest precision (0.539) and F1 score (0.698), and it achieves an outstanding AUC of 0.901. This demonstrates CNSL's superior ability to differentiate between misinformation and non-misinformation spread, especially when the initial condition includes both sources of misinformation and infected agents. The recall of 0.987 in this scenario also indicates that CNSL can identify nearly all instances of misinformation spread.
Overall, the CNSL method outperforms other rule-based and learning-based methods in most metrics across different simulations and seeding strategies in geo-social networks.
\begin{figure}[t]
    \centering
    \includegraphics[width=0.9\columnwidth]{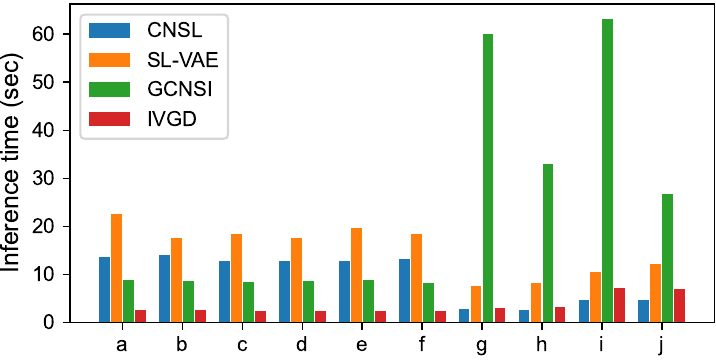}
    \vspace{-4mm}
    \caption{Runtime Comparison with learning based methods for dataset a) LT2LT, b) LT2IC c) LT2SIS, d) IC2LT, e) IC2IC, f) IC2SIS, g) G2S-A-D0, h) G2S-A-D1, i) G2S-B-D0, j) G2S-B-D1}
    \label{fig:runtime_comp}
    \vspace{-3mm}
\end{figure}

\noindent\textbf{Runtime Analysis.}
Figure~\ref{fig:runtime_comp} presents a runtime comparison among four learning-based methods: CNSL, SL-VAE, GCNSI, and IVGD across ten different diffusion configurations (a to j). CNSL, which is our method, shows a competitive inference time in all datasets when compared to the SL-VAE.
In cross-platform communication network datasets (a) LT2LT, b) LT2IC, c) LT2SIS, d) IC2LT, e) IC2IC, and f) IC2SIS)), CNSL demonstrates an inference time that is neither the fastest nor the slowest, indicating a balanced computational demand for these more complex scenarios.
However, in datasets geo-social information spread data (g) G2S-A-D0, h)G2S-A-D1, i)G2S-B-D0, and j)G2S-B-D1), CNSL's runtime is noticeably lower, suggesting that while CNSL is highly effective in identifying misinformation spread. 
Overall, CNSL shows a strength in providing a good balance between accuracy and computational efficiency. While there are scenarios where CNSL's runtime is higher, these may correlate with more complex network conditions where deeper analysis is necessary, which CNSL seems to handle without compromising the quality. This makes CNSL a robust method for practical applications where runtime is a critical factor alongside precision and accuracy.
\begin{figure}[t]
    \centering
    \includegraphics[width=0.9\columnwidth]{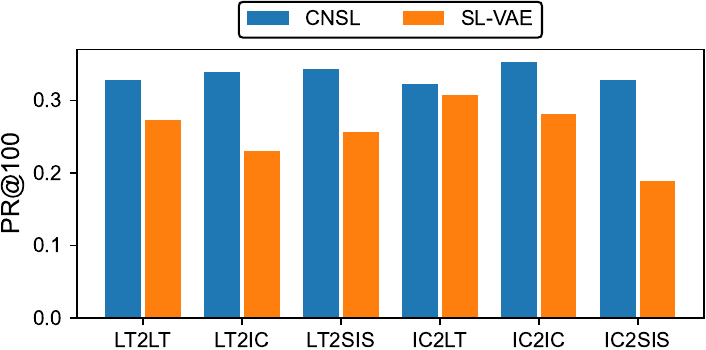}
    \vspace{-3mm}
    \caption{Precision@100: the precision rate of the top 100 nodes being predicted as seed nodes. The comparison is conducted between our method: CNSL and the current state-of-the-art: SL-VAE.}
    \label{fig:pr_comp}
    \vspace{-5mm}
\end{figure}

\noindent\textbf{Precision analysis at top 100 nodes predicted by models.}
Figure~\ref{fig:pr_comp} illustrates the precision at top 100 (PR@100) comparison between CNSL and the state-of-the-art SL-VAE across various diffusion patterns. PR@100 measures the precision rate of the top 100 nodes predicted as seed nodes, indicating how accurately each method can identify the most influential nodes in the spread of information or misinformation. CNSL shows a strong performance in this metric, outperforming SL-VAE in all diffusion patterns. CNSL exhibits higher PR@100 rates, indicating that it is more precise in identifying the key seed nodes. This precision is crucial in scenarios where it is important to quickly and accurately pinpoint the main drivers of information spread within a network. Notably, CNSL's precision suggests that its algorithm is particularly adept at handling complex diffusion patterns where the identification of influential nodes is more challenging.
The strength of CNSL, as highlighted by Figure~\ref{fig:pr_comp}, lies in its ability to consistently rank the most relevant nodes higher than SL-VAE. The precision at the top 100 nodes is essential for practical applications where interventions need to be targeted and efficient, such as in the case of misinformation containment or viral marketing.


\section{Conclusion}
In conclusion, information diffusion source localization on cross-networks requires locating the origins of information diffusion within and across networks. We propose a Cross-Network Source Localization (CNSL) framework in this work, which stands as a pivotal advancement in addressing the complexities introduced by cross-network environments, where traditional source localization methods fall short. By ingeniously approximating the distribution of diffusion sources through mean-field variational inference, encoding both static and dynamic features of nodes via a disentangled generative prior, and uniquely modeling the diffusion dynamics of interconnected networks, CNSL offers a comprehensive solution to the problem. Extensive experiments, including quantitative analysis, case studies, and runtime analysis,  have been conducted to verify the effectiveness of the framework across different real-world and synthetic cross-networks.  The significance of this work lies not only in its methodological innovation but also in its practical implications for safeguarding the integrity and reliability of information in an increasingly interconnected digital world.

	\bibliographystyle{ACM-Reference-Format}
	\bibliography{reference}

\appendix
\section{CNSL Technical Supplements}
\subsection{Derivation of Eq. \eqref{eq:decouple}}
$\log p_{\psi}(y_t|x_s, G_s, G_t) = \log [\sum_{x_t} p_{\psi_1}(x_t|x_s, G_s) \cdot p_{\psi_2}(y_t|x_t, G_t)]$, where $x_t$ inherited infection probability from $y_s$. In practice, we assume $p_{\psi_1}(x_t|x_s, G_s)$ follows delta distribution, where only the $x_t$ is 1 that corresponds to the $x_s$ and the rest of $x_t$’s are 0. This property is also assumed in many works \cite{razavi2018preventing} using VAE. Therefore, $\log p_{\psi}(y_t|x_s, G_s, G_t)$ is simplified as Eq. \eqref{eq:decouple}.

\subsection{Graphical Model of CNSL}
\begin{figure}[t]
    \centering
    \includegraphics[width=0.9\columnwidth]{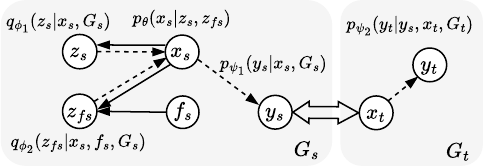}
    \vspace{-3mm}
    \caption{The graphical model for CNSL, where the solid arrows indicate the variational approximation $q_{\phi_1}(z_s|x_s, G_s)$ and $q_{\phi_2}(z_{fs}|x_s,f_s, G_s)$ to the intractable posterior $p(Z|x_s,f_s, G_s)$. Dashed arrows denote the generative process that decodes $x_s$ from $p_{\theta}(x_s|Z)$ and predicts the information diffusion $p_{\psi}(y_t|x_s, \mathcal{G})$. } 
    \label{fig:CNSL_graphical}
    \vspace{-2mm}
\end{figure}

We provide the graphical model for the CNSL framework in Figure \ref{fig:CNSL_graphical}. As shown in the figure, solid arrows indicate the variational approximation $q_{\phi_1}(z_s|x_s, G_s)$ and $q_{\phi_2}(z_{fs}|x_s,f_s, G_s)$ to the intractable posterior $p(Z|x_s,f_s, G_s)$. Dashed arrows denote the generative process that decodes $x_s$ from $p_{\theta}(x_s|Z)$ and predicts the information diffusion $p_{\psi}(y_t|x_s, \mathcal{G})$. The two directional arrow between $y_s$ and $x_t$ indicates $x_t$ inherits the infection probability from the diffusion observation $y_s$ through bridging nodes $L$.

\section{Experiment Supplement}
\subsection{Case Study} \label{subsec:cases}
\begin{figure}[h]
 \centering
		\subfloat[LT2LT]{\label{fig: viz_LT2LT}			\hspace{3mm}\includegraphics[width=0.2\textwidth, trim = 0cm 5cm 0cm 5cm, clip]{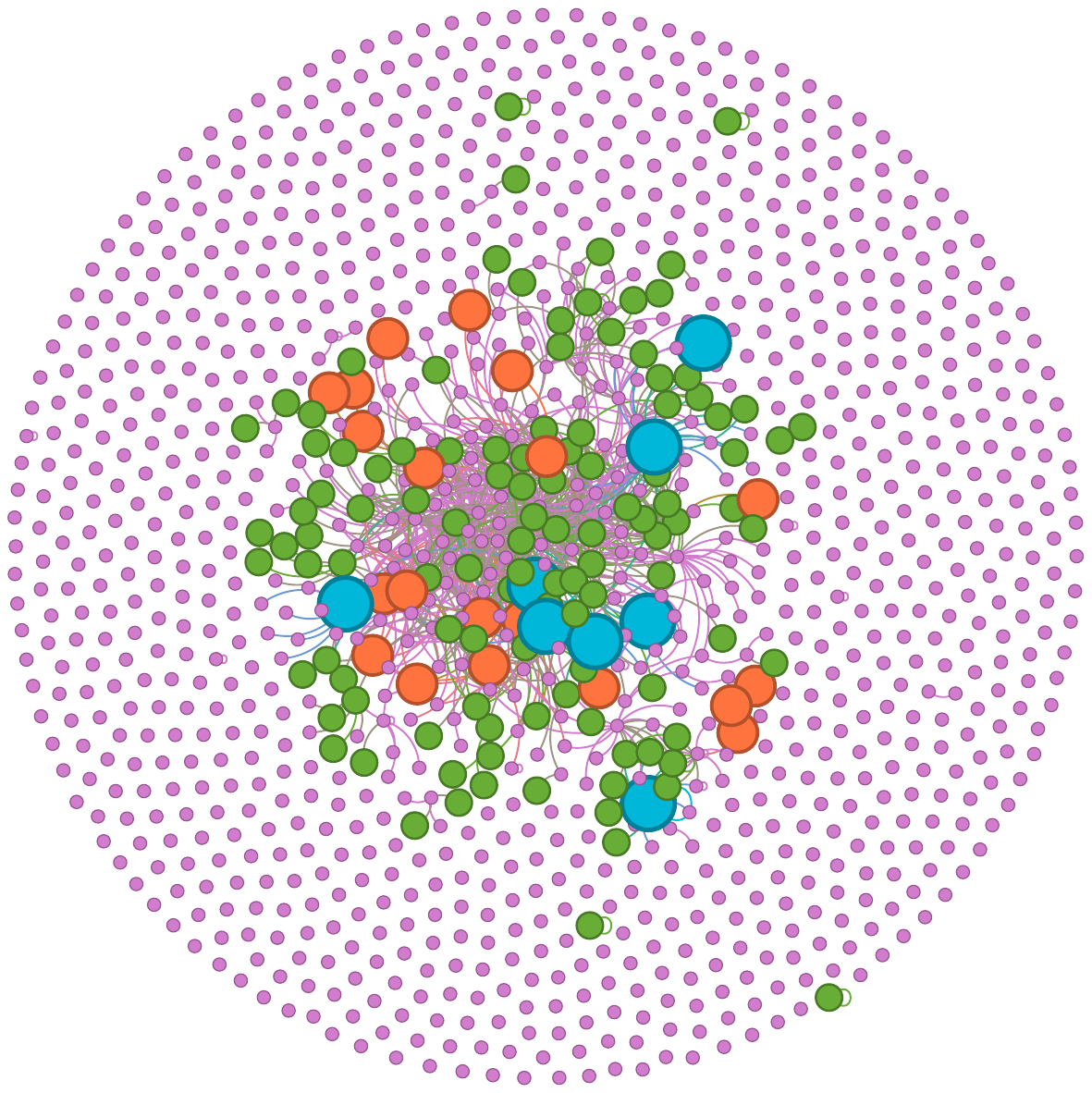}}\\
		\subfloat[LT2IC]{\label{fig: viz_LT2IC}
			\hspace{3mm}\includegraphics[width=0.2\textwidth, trim = 0cm 5cm 0cm 5cm, clip]{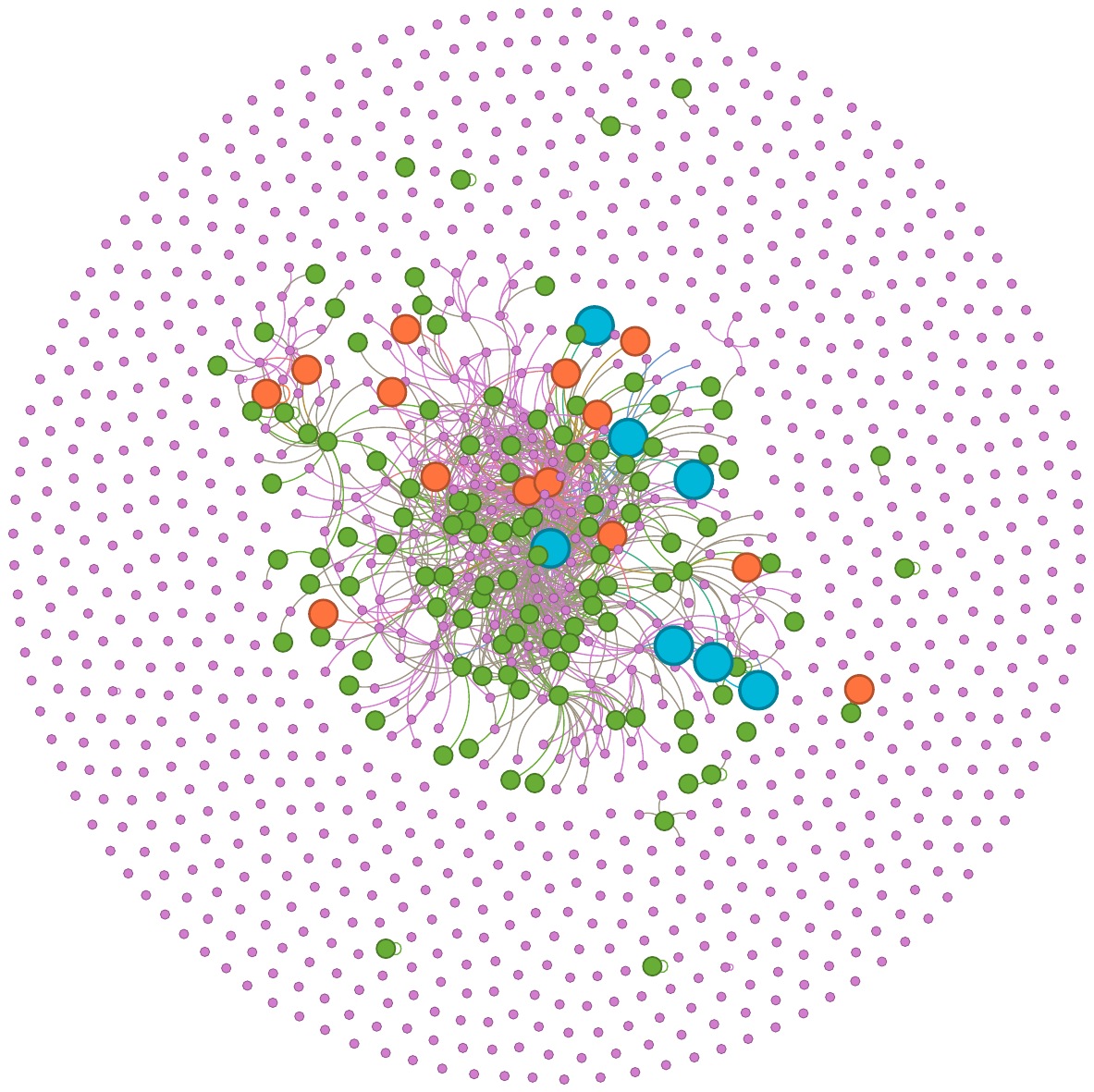}}
		\subfloat[LT2SIS]{\label{fig: viz_LT2SIS}
			\hspace{3mm}\includegraphics[width=0.2\textwidth, trim = 0cm 5cm 0cm 5cm, clip]{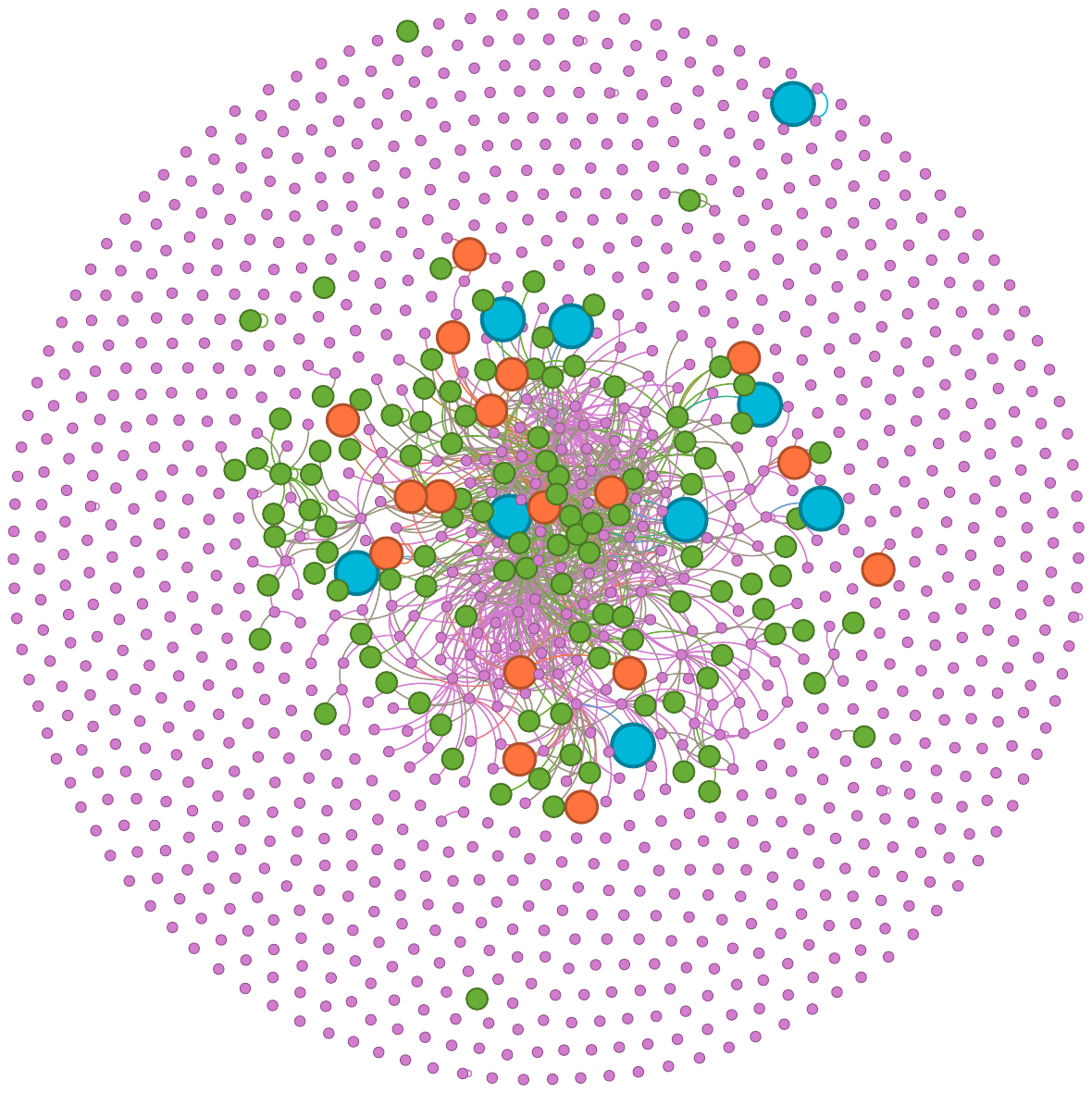}}

		\caption{Seed inference by CNSL.}
		\label{fig: visualization_infer}
 \vspace{-3mm}
	\end{figure}

In a case study depicted in Figure~\ref{fig: visualization_infer}, we illustrate the distribution of selected seed nodes. Here violet nodes represent the nodes that are not seeds. On the other hand, green nodes are the original seeds that were not selected by CNSL; orange nodes are wrongly identified as seeds by CNSL; and the Blue color nodes are correctly identified as seeds by CNSL.

 \pagebreak 
 \subsection{Algorithm}
 \begin{algorithm}
\caption{CNSL Training Framework}
\label{alg:training}
\begin{algorithmic}[1]
\Require $G_s$, $G_t$, $f_s$, $L$, $x_s$, $y_t$
\Ensure Trained $q_{\phi}(\cdot)$, $p_{\theta}(\cdot)$, and $p_{\psi}(\cdot)$
    \For {$epoch$ in $1$ to $num\_epochs$}
        \For {each batch in $train\_set$}
            \State $z_s = q_{\phi_1}(x_s, G_s)$ \Comment{Dynamic Encoder}
            \State $z_{fs} = q_{\phi_2}(x_s, x_{fs}, G_s)$ \Comment{Static Encoder}
            \State $\hat x_s = p_{\theta}(z_s,z_{fs})$ \Comment{Decoder}
            \State $\hat y_s = p_{\psi_1}(\hat x_s, G_s)$  \Comment{Source Network Diffusion}
            \State $\hat x_t \gets \hat y_s$ \Comment{$L = \{(v_s, v_t) | v_s \in V_s, v_t \in V_t\}$}
            \State $\hat y_t = p_{\psi_2}(\hat x_t, G_t)$  \Comment{Target Network Diffusion}
            \State Calculate $\mathcal{L}_{\text{train}}$ \Comment{Equation \eqref{eq:objective}}
            \State Backpropagate loss
            \State Update model parameters 
        \EndFor
    \EndFor
\end{algorithmic}
\end{algorithm}
For training, we want to use observed $x_s$ and $y_t$ to learn the approximate posterior $q_{\phi}(Z|x_s, \mathcal{G})$, the decoding function $p_{\theta}(x_s|Z)$, and the cross-network diffusion prediction function $p_{\psi}(y_t|x_s, \mathcal{G})$. Specifically, we separately obtain two latent variables $z_s$ and $z_{fs}$ in Line $2$-$3$. Both $z_s$ and $z_{fs}$ are fed to reconstruct $\hat x_s$ in Line $5$. After the seed set reconstruction, we conduct cross-network diffusion prediction as shown in Line $6$-$8$. The backpropagation is calculated based on Eq. \eqref{eq:objective} that consists of seed nodes reconstruction error, diffusion estimation error, as well as constraints of KL divergence and influence monotonicity.

 \begin{algorithm}[h]
\caption{CNSL Inference Framework}
\label{alg:inference}
\begin{algorithmic}[1]
\Require $p_{\theta}(x_s|z_s, z_{fs})$; $p_{\psi_1}(y_s|x_s, G_s)$; $p_{\psi_2}(y_t|y_s, x_t, G_t)$; the number of iteration $\eta$; learning rate $\alpha$.
\Ensure $\hat x_s$
\State $\bar z_s = \frac{1}{k} \sum\nolimits^k_i q_{\phi_1}( z_s|\hat x^{(i)}_s, \mathcal{G})$ \Comment{$\hat x^{(i)}_s$ sampled from training set.}
\State $\bar z_{fs} = \frac{1}{k} \sum\nolimits^k_i q_{\phi_2}(z_s|\hat x^{(i)}_s, \mathcal{G})$ \Comment{$\hat x^{(i)}_s$ sampled from training set.}
    \For {$i=0, ..., \eta$}
    \State $\hat x_s = p_{\theta}(\bar z_s,\bar z_{fs})$ \Comment{Decoder}
            \State $\hat y_s = p_{\psi_1}(\hat x_s, G_s)$  \Comment{Source Network Diffusion}
            \State $\hat x_t \gets \hat x_s$ \Comment{$L = \{(v_s, v_t) | v_s \in V_s, v_t \in V_t\}$}
            \State $\hat y_t = p_{\psi_2}(\hat x_t, G_t)$  \Comment{Target Network Diffusion}
        \State $\bar z_s \leftarrow \bar z_s - \alpha\cdot \nabla \mathcal{L}_{\text{pred}}(\hat y_t, \bar z_s, \bar z_{fs})$
    \EndFor
    \State $\hat x_s = p_{\theta}(\bar z_s,\bar z_{fs})$
\end{algorithmic}
\end{algorithm}
For the seed set inference, we first sample $k$ different $\hat x_s^{(i)}$ from the training set, and we marginalize them to obtain two latent variables $\bar z_s$ and $\bar z_{fs}$ (Line $1$-$2$). For $\eta$ iterations, we decode the predicted $\hat x_s$ based on $(\bar z_s, \bar z_{fs})$ (Line $4$) and conduct cross-network information diffusion prediction (Line $5$-$7$). The error between predicted $\hat y_t$ and the observed $y_t$ is leveraged to update $\bar z_s$ based on Eq. \eqref{eq: infer_2}.

\end{document}